\documentclass[12pt]{iopart}

\usepackage{graphicx}
\usepackage{iopams}
\begin{document}

\title{Evolutionary game dynamics in inhomogeneous populations}

\author{Xiaojie Chen$^{1,2}$, Feng Fu$^{1,2}$, Long Wang$^{1,2}$ and Tianguang Chu$^{1,2}$ },
\address{$^1$Intelligent Control Laboratory, Center for Systems and Control, Department of Mechanics and Space Technologies, College of Engineering, Peking University, Beijing 100871, China}

\address{$^2$Department of Industrial Engineering and Management, College of Engineering, Peking University, Beijing 100871, China}

\ead{\mailto{longwang@pku.edu.cn}, \mailto{xjchen@pku.edu.cn}}
\begin{abstract}
To our knowledge, the populations are generally assumed to be
homogeneous in the traditional approach to evolutionary game
dynamics. Here, we focus on the inhomogeneous populations. A
simple model which can describe the inhomogeneity of the
populations and a microscopic process which is similar to Moran
Process are presented. By studying the replicator dynamics, it is
shown that this model also keeps the fixed points unchanged and
can affect the speed of converging to the equilibrium state. The
fixation probability and the fixation time of this model are
computed and discussed. In the inhomogeneous populations, there
are different situations that characterize the time scale of
evolution; and in each situation, there exists an optimum solution
for the time to the equilibrium points, respectively. Moreover,
these results on the speed of evolution are valid for infinite and
finite populations.

\end{abstract}

\submitto{\JPA}
\maketitle

\section{Introduction}

Evolutionary game theory has been successfully founded and applied
to the study of biology, economics, and social sciences by Maynard
Smith \cite{1}. Originally, evolutionary game theory was
formulated in terms of infinite populations and the corresponding
replicator dynamics. Consider two strategies A and B in a
population engaged in a game with payoff matrix
\begin{eqnarray*}
\begin{array}{cccccccc}
&&     A&&          B     \\
A&&     a&&           ~b .\\
B&&     c&&           d   \\
\end{array}
\end{eqnarray*}
A typical assumption is that individuals meet each other at random
in infinitely large, well-mixed populations. The fitness (or
payoff) of A and B players are respectively given by
\begin{equation}
\begin{array}{cccccc}
f_A=ax+b(1-x), \\  f_B=cx+d(1-x),
\end{array}
\end{equation}
where $x$ is the frequency of A players and $1-x$ is the frequency
of B players. The average fitness of the population is
\begin{equation}
\bar{f}=xf_A+(1-x)f_B .
\end{equation}
The standard replicator equation which describes evolutionary
dynamics in a infinite population takes the form \cite{2,3}
\begin{eqnarray}
\dot{x} &=&x(f_A-\bar{f}) \nonumber \\
        &=&x(1-x)[(a-b-c+d)x-(d-b)].\label{eq3}
\end{eqnarray}
The equilibrium points are either on the boundary or in the
interior. There are four generic outcomes \cite{4,5,6}: \\
(1) If $a>c$ and $b>d$ then A dominates B; the only stable
equilibrium is $x=1$.\\
(2) If $a<c$ and $b<d$ then B dominates A; the only stable
equilibrium is $x=0$.\\
(3) If $a>c$ and $b<d$ then A and B are bi-stable; both $x=0$ and
$x=1$ are stable equilibria; there is an unstable equilibrium at
$x=(d-b)/(a-b-c+d)$.\\
(4) If $a<c$ and $b>d$ then A and B co-exist; both $x=0$ and $x=1$
are unstable equilibria; the only stable equilibrium is given by
$x=(d-b)/(a-b-c+d)$.

The standard replicator dynamics hold in the limit of infinite
population size. In fact, any real population has finite size and
also computer simulations in structured or unstructured
populations always deal with finite populations \cite{7,8,9,10}.
Therefore, it is natural to study evolutionary game dynamics in
finite populations. In most approaches for finite population size,
each individual interacts with each other individual in the
well-mixed, homogeneous populations. Moreover, stochastic
processes have been introduced to study evolutionary dynamics in
finite populations. Recently, in unstructured finite populations
different mechanisms are applied to study game dynamics, such as
Moran Process, Pairwise Comparison Process, Wright-Fisher Process,
local information, mutation, discounting and active linking
\cite{11,12,13,14,15,16,17}.

To our best knowledge, in the aforementioned approaches to
evolutionary game theory, they are all based on the simplifying
assumptions that the populations are homogeneous and each
individual, which is engaged in symmetric game, is identical to
strategy update. In fact, biological agents in many real
populations are non-identical to their abilities to competition,
survival and reproduction. For instance, the difference in sex,
male or female, plays a significant role in group dominance. The
age, old or young; the strength, strong or weak, etc, are also
factors affecting the individuals' competition and cooperation.
Thus, we here relax the simplifying assumptions and consider that
the populations are inhomogeneous. In our scenario, we aim to
investigate the inhomogeneity's effect in evolutionary game
dynamics. The remainder of this paper is organized as follows: A
simple model is constructed to describe the inhomogeneity of the
populations and a stochastic process for evolutionary game theory
is introduced in \sref{section2}. And then analytical results and
corresponding simulations of the model are provided in
\sref{section3}. Finally, conclusions are made in \sref{section4}.

\section{The Model}\label{section2}

In this model, the populations are well-mixed and each player
interacts with each other player. To describe the inhomogeneity of
the populations, we just assume that two types of players are
distributed randomly in the populations (just like male and female
individuals in a population) \cite{18,19}. For simplicity, we use
E to denote one type players and F to denote the other type
players. Every player has only one type and their distribution is
fixed later on. The concentration of players E and F are denoted
by $v (0\leq v \leq1)$ and $1-v$. All individuals just follow A or
B strategies no matter what types they are. And when players E
interact with other players, the payoff of players E will be
strengthened no matter what strategies players E follow; while the
payoff of players F will keep unchanged no matter what strategies
players F follow when players F interact with other players. Now,
suppose the population consists of $N$ players. The number of
players using strategy A is given by $i$, the number of players
using strategy B is given by $N-i$. If every player interacts with
every other player, the average payoff of A and B are respectively
given from a mean-field theory
\begin{equation}
\begin{array}{ccccc}
\Pi^A_i=\frac{a(i-1)+b(N-i)}{N-1}[vq_1+(1-v)],\\
\Pi^B_i=\frac{ci+d(N-i-1)}{N-1}[vq_1+(1-v)],
\end{array}
\end{equation}
where the parameter $q_1(q_1>1)$ characterizes the rates of
increased payoff of players E. Therefore, the average payoff of
the population at the state is given
\begin{equation}
<\Pi>=[i\Pi^A_i+(N-i)\Pi^B_i]/N.
\end{equation}
Then, the average fitness of strategies A and B are respectively
given by \cite{20}
\begin{equation}
\begin{array}{cccccc}
f_A=1-w+w\Pi^A_i, \\
f_B=1-w+w\Pi^B_i,
\end{array}
\end{equation}
where $w$ measures the intensity of selection. Strong selection
means $w=1$; weak selection means $w\ll1$.

We now describe the selection mechanism process as follows: In
each time step, an individual is chosen with a probability
proportional to its fitness; a second individual is selected
randomly. Then the second individual switches to the first one's
strategy. Moreover, if the second individual is a player E, it
will weaken the probability to switch to the first one's strategy;
otherwise, it will keep the probability to switch to the first
one's strategy. And we write the probability that the number of A
individuals increases from $i$ to $i+1$ as
\begin{eqnarray}
T^+(i) &=&\frac{if_A}{if_A+(N-i)f_B}\cdot\frac{N-i}{N}\cdot[vq_2+(1-v)] \nonumber\\
        &=&\frac{1-w+w\Pi^A}{1-w+w<\Pi>}\cdot\frac{i}{N}\cdot\frac{N-i}{N}\cdot[vq_2+(1-v)],
\end{eqnarray}
where the parameter $q_2$ characterizes the strength of reduced
switching activity if the second individual is occupied by an
individual of type E. Since players E can strengthen their payoff,
they are not sensitive to switch their strategies, therefore, we
set $q_2<1$ . The probability that the number of A individuals
decreases from $i$ to $i-1$ is
\begin{eqnarray}
T^-(i)=\frac{1-w+w\Pi^B}{1-w+w<\Pi>}\cdot\frac{i}{N}\cdot\frac{N-i}{N}\cdot[vq_2+(1-v)].
\end{eqnarray}
Consequently, the probability that the number of A individuals
remains constant is $T(i)=1-T^+(i)-T^-(i)$. Since $T^-(N)=0$ and
$T^+(0)=0$, this process has absorbing states at $i=0$ and $i=N$.
For large populations, a Langevin equation can approximately
describe this process \cite{11}
\begin{eqnarray*}
\begin{array}{cccccc}
\dot{x}=a(x)+b(x)\varepsilon, \\
a(x)=T^+(x)-T^-(x),\\
b(x)=\sqrt{[T^+(x)+T^-(x)]/N},
\end{array}
\end{eqnarray*}
where $x=\frac{i}{N}$ is the fraction of A, $a(x)$ is the drift
term, $b(x)$ is the diffusion term and $\varepsilon$ is
uncorrelated Gaussian noise. For large $N$, $b(x)$ vanishes with
$1/\sqrt{N}$, this equation becomes
\begin{eqnarray}
\dot{x} &=&x(1-x)\cdot\frac{w[\Pi^A(x)-\Pi^B(x)]}{1-w+w<\Pi(x)>}\cdot[vq_2+(1-v)] \nonumber\\
        & \triangleq &F(v)\cdot
        x(x-1)[(b+c-a-d)x+(d-b)]/G(w,x,v)\label{eq9},
\end{eqnarray}
where
\begin{eqnarray*}
F(v) &=&[vq_1+(1-v][vq_2+(1-v)]\nonumber\\
     &=&(q_1-1)(q_2-1)v^2+(q_1+q_2-2)v+1,\label{eq14}
\end{eqnarray*}
\begin{eqnarray*}
G(v,x,w) & = &\frac{1-w}{w}+[vq_1+(1-v)]\cdot\nonumber \\
         &   &[(a-b-c+d)x^2+(b+c-2d)x+d].\label{eq15}
\end{eqnarray*}
\eref{eq9} is the replicator dynamics equation for this model. For
$v=0$, the replicator dynamics equation for the Moran Process in
homogeneous populations is recovered. For $0<v<1$, inhomogeneity
is introduced in the system as there are two types of players in
populations. Subsequently, the replicator dynamics, the fixation
probability and the fixation time of this model are to be
investigated and discussed for different values of the parameters.

\section{Analytical Results and Corresponding Simulations}\label{section3}

For the Moran Process in homogeneous populations,
\begin{equation}
\dot{x}=x(x-1)[(b+c-a-d)x+(d-b)]/\Gamma(x,w),\label{eq10}
\end{equation}
where $\Gamma(x,w)=\frac{1-w}{w}+[(a-b-c+d)x^2+(b+c-2d)x+d]$.
Comparing with (\ref{eq3}) and (\ref{eq10}), \eref{eq9} also has
the three same equilibria: $x=0$, $1$ and $(d-b)/(a-b-c+d)$ and
keeps the fixed points unchanged. Moreover, there are apparently
four same generic cases for the stable equilibrium points to
\eref{eq3} by studying \eref{eq9}. To illustrate this, let us
consider an example. Consider the payoff matrix
\begin{eqnarray*}
\begin{array}{cccccccc}
&&     A&&           B     \\
A&&     0.3&&        0.5 ~.  \\
B&&     0.1&&        0.2   \\
\label{eq1}\end{array}
\end{eqnarray*}
\begin{figure}
\centering
\includegraphics[scale=1]{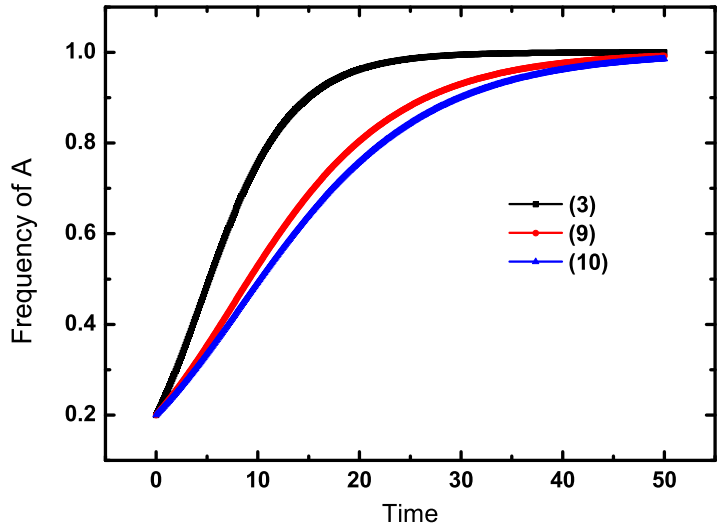}
\caption{Frequency of A as a function of time for different
equations from an initial state with $20\%$ A, given fixed values
of $q_1=2$, $q_2=0.4$ and $w=0.4$ for (\ref{eq9}) and
(\ref{eq10}).\label{fig1}}
\end{figure}

The fitness of A is greater than the fitness of B in this example.
Hence, we say that A dominates B. \Fref{fig1} shows the evolution
from a state with $20\%$ A into the end state with all A. Since
$a=0.3>c=0.1$ and $b=0.5>d=0.2$, \fref{fig1} confirms the
theoretical predictions.

In fact, the differences among the three dynamics equations amount
to a dynamics rescaling of time. And $F(v)$ and $G(v,x,w)$ in
(\ref{eq9}) are factors influencing the time scale only. They
would affect only the speed of evolution, but would not influence
the long-run behavior. Then, we would like to show that how they
affect the time scale for different values of the parameters. In
this model for the fixed values of $q_1$, $q_2$ and $w$,
$G(v,x,w)\approx\frac{1-w}{w}$ is constant with weak selection,
then only $F(v)$ can influence the time scale for the variable
$v$. Here, $F(v)$ has a maximum at $v=v_c$ for different $v\in[0
~1]$, and $F(v)=(q_1-1)(q_2-1)v^2+(q_1+q_2-2)v+1\leq F(v_c)$, then
there exists the optimum $v_c$ to converge fastest to the
equilibrium state. Since $(q_1-1)(q_2-1)<0~(q_1>1, q_2<1)$, there
are three cases for different relationships between $q_1$ and
$q_2$:
\begin{eqnarray*}
F(v)&=&(q_1-1)(q_2-1)[v+\frac{q_1+q_2-2}{2(q_1-1)(q_2-1)}]^2+\nonumber\\
    &&1-\frac{(q_1+q_2-2)^2}{4(q_1-1)(q_2-1)}.
\end{eqnarray*}
(1) If $q_1+q_2<2$, then $F(v)$ has its maximum at $v_c=0$.\\
(2) If $2\leq q_1+q_2<2q_1q_2$, then $F(v$ has its maximum at
$v_c=1$.\\
(3) If $q_1+q_2\geq2q_1q_2$, then $F(v)$ has its maximum at
$v_c=\frac{2-q_1-q_2}{2(q_1-1)(q_2-1)}$.\\
Especially, the interesting relationship between $q_1$ and $q_2$
for $q_1>1$ and $q_2<1$ is $q_1q_2=1$. In this case, then
$q_1+q_2>2q_1q_2=2$ and there is only one outcome for $v_c$:
$v_c=\frac{2-q_1-q_2}{2(q_1-1)(q_2-1)}=0.5$.

The four outcome predictions, which can respectively reflect the
four relationships between $q_1$ and $q_2$, are found from the
replicator dynamics equation, thus they are justified for infinite
or large finite populations. In other words, it can converge
fastest to the equilibrium state when $v=v_c$ for infinite
populations. However, in finite populations, if the mean time to
fixation becomes very large, the model may be limited interest,
therefore, discussion on the fixation time $T_v$ is an interesting
topic. Here, whether the fixation time $T_v$ in finite populations
has a minimum at $v_c$ respectively corresponding to the four
situations in infinite populations is a more interesting topic.
Indeed, the four outcome predictions for infinite populations are
still valid for small finite populations. For finite populations,
$T_v$ means that the time from an initial state $x_0$ to the
equilibrium state and can be calculated by \cite{13,21}
\begin{equation}
T_v=N\int_0^1t(x,x_0)dx,\label{eq11}
\end{equation}
where
$$t(x,x_0)=\frac{2[S(1)-S(x_0)]S^2(x)}{b^2(x)S(1)S(x_0)}\cdot\exp{[\int_0^x\frac{2a(y)}{b^2(y)}dy]},
(0\leq x \leq x_0)$$
$$t(x,x_0)=\frac{2[S(1)-S(x)]S(x)}{b^2(x)S(1)}\cdot\exp{[\int_0^x\frac{2a(y)}{b^2(y)}dy]},
(x_0\leq x \leq 1)$$ \\
and
$$S(x)=\int_0^x\exp{[-\int_0^y\frac{2a(z)}{b^2(z)}dz]}dy.$$

\begin{figure}
\centering
\includegraphics[scale=1]{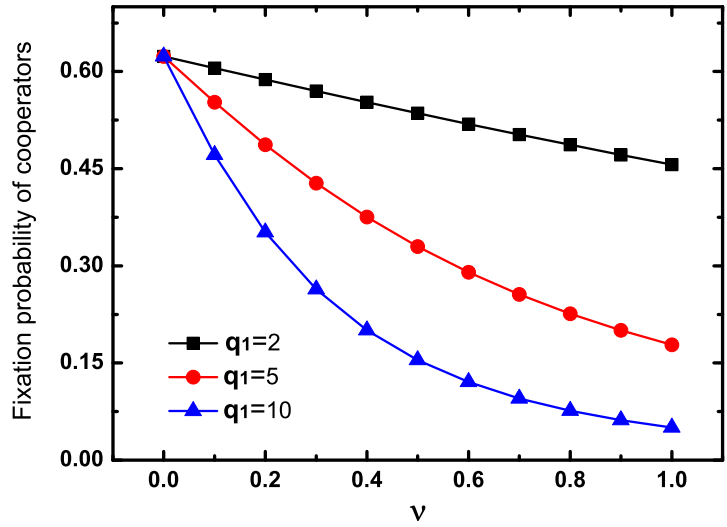}
\caption{Fixation probability of $k=80$ cooperators in a
Prisoner's Dilemma as a function of $v$ for different rates $q_1$,
given a fixed value of $N=100$.\label{fig2}}
\end{figure}
\begin{figure}
\centering
\includegraphics[scale=0.8]{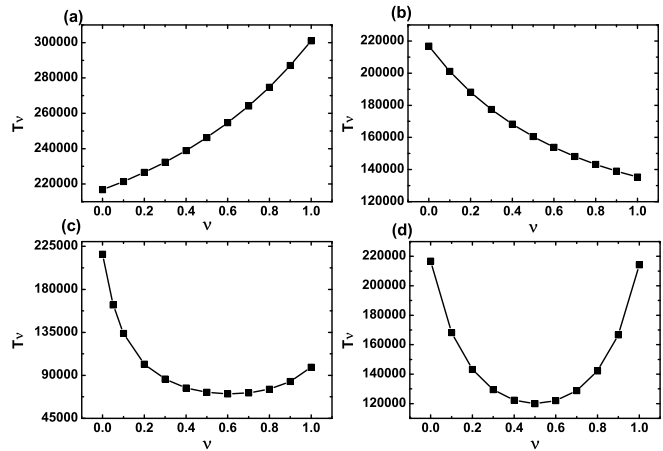}
\caption{The fixation time $T_v$ as a function of $v$ for
$N=1000$, given different relationships between $q_1$ and $q_2$:
(a). $q_1=1.2$ and $q_2=0.6$; in this case, $v_c=0$. (b). $q_1=2$
and $q_2=0.8$; in this case, $v_c=1$. (c). $q_1=8.5$ and
$q_2=0.25$; in this case, $v_c=0.6$. (d). $q_1=5$ and $q_2=0.2$;
in this case, $v_c=0.5$\label{fig3}}
\end{figure}
As this can be done numerically in general, the corresponding
simulation results are shown below. Before computing the fixation
time, let us first investigate the fixation probability for this
model. The fixation probability $\phi_k$ with $k$ players using
strategy A is given by \cite{22}
\begin{eqnarray}
\phi_k &=&\frac{1+\sum_{j=1}^{k-1}\prod_{i=1}^j
\frac{T^-(i)}{T^+(i)}}{1+\sum_{j=1}^{N-1}\prod_{i=1}^j
\frac{T^-(i)}{T^+(i)}}\label{eq12}\\
      &=&\frac{\displaystyle 1+\sum_{j=1}^{k-1}\prod_{i=1}^j
\frac{(1-w)(N-1)+w(vq_1+1-v)[ci+d(N-i-1)]}{(1-w)(N-1)+w(vq_1+1-v)[a(i-1)+b(N-i)]}}{\displaystyle
1+\sum_{j=1}^{N-1}\prod_{i=1}^j
\frac{(1-w)(N-1)+w(vq_1+1-v)[ci+d(N-i-1)]}{(1-w)(N-1)+w(vq_1+1-v)[a(i-1)+b(N-i)]}}.
\nonumber
\end{eqnarray}

Now let us take the Prisoner's Dilemma for example. In most
papers, the Prisoner's Dilemma is determined by the payoff matrix
\begin{eqnarray*}
\begin{array}{cccccccc}
&&&    C&&          D     \\
C&&&     b-c&&         -c   .\\
D&&&     b&&           0   \\
\end{array}
\end{eqnarray*}
To assure that the fitness of C and D are always positive, the
payoff matrix becomes
\begin{eqnarray*}
\begin{array}{cccccccc}
&&&    C&&           D     \\
C&&&     b-c&&         -c   \\
D&&&     b&&           0   \\
\end{array}
~~~\Rightarrow~~~
\begin{array}{cccccccc}
&&&     C&&           D     \\
C&&&     b&&         0   .\\
D&&&     b+c&&           c   \\
\end{array}
\end{eqnarray*}

In the following simulation results, the initial frequency of
cooperators is $80\%$, and we set $b=3$, $c=2$ and $w=0.0001$. In
\fref{fig2}, we show the fixation probability of a Prisoner's
Dilemma starting with $80\%$ cooperators. Clearly, cooperators are
always dominated by defectors. It shows that stronger rates $q_1$
decrease the fixation probability of cooperators and the fixation
probability of cooperators monotonically decreases when the value
of the parameter $v$ increases with a given fixed $q_1$. These
results can be understood in the following way. When the values of
$q_1$ or $v$ increase, it results in that the temperature of
selection is enhanced. For the Prisoner's Dilemma, the average
payoff of cooperators is less than the average payoff of
defectors. Therefore, the fixation probability of cooperators
decreases for the Prisoner's Dilemma when the temperature of
selection is increased \cite{19}. Moreover, we have found that the
fixation probability has nothing to do with the strength of
switching activity $q_2$ from (\ref{eq12}). In \fref{fig3}, we
show the fixation time of a Prisoner's Dilemma starting with
$80\%$ cooperators for $N=1000$. The fixation time from
(\ref{eq11}) for different situations are computed, respectively.
In \fref{fig3}(a), $q_1+q_2=1.8<2$. In this case, $T_v$ has its
minimum at $v=0$. And we observe that $v_c=0$ from \fref{fig3}(a).
In \fref{fig3}(b), $2<q_1+q_2=2.8<3.2=2q_1q_2$. In this case,
$T_v$ has its minimum at $v=1$. And we observe that $v_c=1$ from
\fref{fig3}(b). In \fref{fig3}(c), $2q_1q_2=4.25<q_1+q_2=8.75$, In
this case, $T_v$ has its minimum at
$v=\frac{2-8.5-0.25}{2\times(8.5-10\times(0.25-1)}=0.6$. And we
observe that $v_c=0.6$ from \fref{fig3}(c). In \fref{fig3}(d),
$2<q_1+q_2=5.2$ and $q_1q_2=1$. In this case, $T_v$ has its
minimum at $v=0.5$. And we observe that $v_c=0.5$ from
\fref{fig3}(d). These results for finite populations are totally
in very good agreement with theoretical predictions for infinite
populations and these figures confirm that the fixation time $T_v$
also has its minimum at $v_c$. Moreover, these results for the
fixation time in finite populations are still valid even if $w$
does not satisfy the condition: $w\ll1$.

\section{Conclusions}\label{section4}

To sum up, we have studied the evolutionary game dynamics in
inhomogeneous populations. We have provided  a model by
description of a microscopic process which is similar to Moran
Process. Comparing with standard replicator and Moran Process
dynamics, it also keeps the fixed points unchanged. Nevertheless,
this can affect the speed of converging to the equilibrium state.
We have also calculated the fixation probability and the fixation
time, and found that there exists an optimum solution to converge
fastest to the stable equilibria. This result requires no limiting
assumption on population size. As is known, how to decrease the
mean time to the fixed state from an initial state is an important
quantity \cite{23}. From this perspective, our results on
inhomogeneous populations may shed light on this issue.

\ack{Discussions with Jing Wang, Zhuozheng Li and Zoujin Ouyang
are gratefully acknowledged. This work was supported by National
Natural Science Foundation of China (NSFC) under grant No.
60674050 and No. 60528007, National 973 Program (Grant No.
2002CB312200), National 863 Program (Grant No. 2006AA04Z258) and
11-5 project (Grant No. A2120061303).}


\section*{References}
\begin{thebibliography}{23}
\bibitem{1}
J.~M.~Smith 1982 {\it Evolution and the Theory of Games} (London:
Cambridge University Press)
\bibitem{2}
P.~D.~Taylor and L.~Jonker 1978 {\it Math. Biosci.} {\bf 40} 145
\bibitem{3}
J.~Hofbauer and K.~Sigmund 1998 {\it Evolutionary Games and
Population Dynamics} (London: Cambridge University Press)
\bibitem{4}
C.~Taylor, D.~Fudenberg, A.~Sasaki and M.~A.~Nowak 2004 {\it Bull.
Math. Biol.} {\bf 66} 1621
\bibitem{5}
L.~A.~Imhof and M.~A.~Nowak 2006 {\it J. Math. Biol.} {\bf 52} 667
\bibitem{6}
C.~Taylor and M.~A.~Nowak 2006 {\it Theor. Popul. Biol.} {\bf 69}
243
\bibitem{7}
C.~Hauert and M.~Doebeli 2004 {\it Nature} {\bf 428} 643
\bibitem{8}
H.~Ohtsuki, C.~Hauert, E.~Lieberman and M.~A.~Nowak 2006 {\it
Nature} {\bf 441} 502
\bibitem{9}
F.~C.~Santos and J.~M.~Pacheco 2005 \PRL {\bf 95} 098104
\bibitem{10}
J.~Vukov, G.~Szab\'{o} and A.~Szolnoki 2006 {\it Phys. Rev.} E
{\bf 73} 067103
\bibitem{11}
A.~Traulsen, J.~C.~Claussen and C.~Hauert 2005 \PRL {\bf 95}
238701
\bibitem{12}
A.~Traulsen, M.~A.~Nowak and J.~M.~Pacheco 2006{\it Phys. Rev.} E
{\bf 74} 011909
\bibitem{13}
A.~Traulsen, J.~M.~Pacheco and L.~A.~Imhof 2006 {\it Phys. Rev.} E
{\bf 74} 021905
\bibitem{14}
C.~Hauert, F.~Michor, M.~A.~Nowak and M.~Doebeli 2006 {\it J.
Theor. Biol.} {\bf 239} 195
\bibitem{15}
D.~Fudenberg, M.~A.~Nowak, C.~Hauert and L.~A.~Imhof 2006 {\it
Theor. Popul. Biol.} {\bf 70} 262
\bibitem{16}
M.~Willensdorfer and M.~A.~Nowak 2005 {\it J. Theor. Biol.} {\bf
237} 355
\bibitem{17}
C.~P.~Roca, J.~A.~Cuseta and A.~S\'{a}nchez 2006 \PRL {\bf97}
158801
\bibitem{18}
A.~Szolnoki and G.~Szab\'{o} 2006 {\it Preprint} q-bio.PE/0610001
\bibitem{19}
A.~Traulsen, M.~A.~Nowak and J.~M.~Pacheco 2007 {\it J. Theor.
Biol.} {\bf 244} 349
\bibitem{20}
M.~A.~Nowak, A.~Sasaki, C.~Taylor and D.~Fudenberg 2004 {\it
Nature} {\bf 428} 646
\bibitem{21}
W.~J.~Ewens 1979 {\it Mathematical Population Genetics} (Berlin:
Springer)
\bibitem{22}
S.~Karlin and H.~M.~A.~Taylor 1975 {\it A first course in
stochastic process} (New York: Academic press)
\bibitem{23}
C.~Taylor, Y.~Iwasa and M.~A.~Nowak 2006 {\it J. Theor. Biol.}
{\bf 243} 245
\end{thebibliography}
\end{document}